\documentclass[preprint,showpacs,tightenlines,preprintnumbers,prc,amsmath,amssymb]{revtex4}
\usepackage{graphicx}% Include figure files
\usepackage{epsfig}
\usepackage{dcolumn}% Align table columns on decimal point
\usepackage{bm}% bold math
\def\lsim{\mathrel{\rlap{\lower4pt\hbox{\hskip1pt$\sim$}}
    \raise1pt\hbox{$<$}}}         %less than or approx. symbol
\def\gsim{\mathrel {\rlap{\lower4pt\hbox{\hskip1pt$\sim$}}
    \raise1pt\hbox{$>$}}}         %greater than or approx. symbol

\def\re{\mbox{Re}}

\newcommand{\mpi}{m _\pi}

\begin{document}

\title{Chiral potentials, perturbation theory, and the ${}^1$S$_0$
  channel of $NN$ scattering}
% Force line breaks with \\

\author{Deepshikha Shukla$^1$}\email{choudhur@gwu.edu}
\author{Daniel R. Phillips$^2$}\email{phillips@phy.ohiou.edu}
\author{Eric Mortenson$^2$}
\affiliation{$^1$Department of Physics, George Washington
University, Washington DC 20052. \\
$^2$Department of Physics and Astronomy, Ohio University, Athens, OH
45701.}

\date{\today}

\begin{abstract}
We use nucleon-nucleon phase shifts obtained from experimental data,
together with the chiral expansion for the long-distance part of the
$NN$ interaction, to obtain information about the short-distance piece
of the $NN$ potential that is at work in the ${}^1$S$_0$ channel.  We
find that if the scale $R$ that defines the separation between
``long-'' and ``short-'' distance is chosen to be $\lsim 1.8$ fm then
the energy dependence produced by short-distance dynamics is well
approximated by a two-term polynomial for $T_{\rm lab} \leq 200$ MeV.
We also find that a quantitative description of $NN$ dynamics is
possible, at least in this channel, if one treats the long-distance
parts of the chiral $NN$ potential in perturbation theory.  However,
in order to achieve this we have to choose a separation scale $R$ that is
larger than 1.0 fm.
\end{abstract}

\maketitle

\section{Introduction}

\label{sec-intro}

The chiral symmetry of the strong interaction places significant
constraints on hadron-hadron interactions at low energies. Chiral
perturbation theory ($\chi$PT) implements these constraints in a
systematic fashion. (For recent reviews see Refs.~\cite{Be05,SS05}.)
In contrast to the situation in the single-nucleon and
mesonic sectors, low-energy nucleon-nucleon interactions do not vanish in the
chiral limit, with experimental manifestations of the strength of the $NN$
interaction at low energies being provided by the large $np$ scattering
length in the ${}^1$S$_0$ channel and the presence of a bound state
(the deuteron) in the ${}^3$S$_1$ channel.

This complicates the application of perturbation-theory methods to
multi-nucleon interactions. Weinberg's pioneering efforts~\cite{wein}
in the early nineties proposed surmounting this difficulty by making
a $\chi$PT expansion for the $NN$ potential, $V$, that goes into the
Schr\"odinger equation, i.e. solving:
\begin{equation}
(E - H_0)|\psi \rangle=V |\psi \rangle,
\label{eq:SE}
\end{equation}
with
\begin{equation}
V=V^{(0)} + V^{(2)} + V^{(3)} + \ldots. \label{eq:V}
\end{equation}
Here the superscripts indicate the power of the (presumably) small
quantities $\frac{m_\pi}{\Lambda_{\chi SB}}$, $\frac{p}{\Lambda_{\chi
    SB}}$ ($p$ is the momentum of the $NN$ collision and
$\Lambda_{\chi SB}$ the scale of chiral-symmetry breaking) that is
present in that piece of $V$. $\chi$PT calculation then reveals that
$V^{(0)}=C_0 \delta^{(3)}(r) + V_{1\pi}$ includes one-pion exchange
and a short-range interaction, $V^{(2)}$ includes higher-derivative
short-range interactions, together with two-pion exchange diagrams
constructed from the leading-order $\chi$PT Lagrangian, and $V^{(3)}$
involves the so-called ``next-to-leading order'' two-pion exchange,
where the $\pi \pi NN$ vertices are those from ${\mathcal L}_{\pi
  N}^{(2)}$ of $\chi$PT. In Weinberg's original paper
there was no distinction made between the power counting for the piece
of the potential that is operative at long distances ($r \sim
1/m_\pi$) and that for the delta functions and their derivatives which
represent the shorter-range mechanisms ($r < 1/m_\pi$) in this
approach. Both are assumed to give a contribution to the $NN$
potential of an order given by naive dimensional analysis in powers of
$p$ and $m_\pi$.

The implications of the expansion (\ref{eq:V}) for $NN$ scattering
data were first examined in detail by Ordo\~nez et al. in their
landmark 1996 paper~\citep{Or96}. These authors analyzed $NN$
scattering, not only in the $^1$S$_0$, but also in a number of other
partial waves. Improvements on this analysis were made in the work of
Epelbaum and co-authors~\citep{Ep99}, while peripheral $NN$ partial
waves were analyzed in first- and second-Born approximation in
Ref.~\citep{Ka97}.  Two sets of authors have subsequently extended
these analyses to fourth order in $V$~\cite{EM02,EM03,Ep05}.  In all
these studies the expansion (\ref{eq:V}) yielded reasonably
convergent results for $NN$ phase shifts.

Kaplan, Savage, and Wise (KSW)~\citep{ksw1, ksw2} have demonstrated
that the assumption that naive dimensional analysis sets the size of
the short-distance pieces of the potential is chirally inconsistent,
in the sense that iterations of the leading-order $V$ (\ref{eq:V}) via
the Schr\"odinger equation (\ref{eq:SE}) generate divergences
proportional to $m_\pi^2$~\cite{Ka96}. But, at leading
order in Weinberg's expansion, there is no counterterm to absorb this
divergence. KSW proposed an alternative expansion, wherein the
delta-function interaction $C_0 \delta^{(3)}(r)$ was promoted to
leading-order, but one-pion exchange retained its
naive-dimensional-analysis scaling, making it sub-leading in this
expansion. This expansion was carefully examined in the $^1$S$_0$
channel~\citep{ksw1,ksw2,Fl00}, and appears to converge reasonably
well there. However, the KSW expansion does not converge in the
$^3$S$_1$ channel for momenta $\gtrsim 100$~MeV~\cite{Fl00}. A
compromise proposal is to expand one-pion exchange about its chiral
limit value, in which case it is leading in the $^3$S$_1$ and
sub-leading in $^1$S$_0$~\cite{Be01}. The resulting series has been
shown to converge, albeit slowly.

However, two-pion-exchange corrections are yet to be considered in
light of the analysis of the $^1$S$_0$ of Refs.~\cite{Ka96,Be01}.  The
discussions in these works focused on establishing the correct power
counting for the $NN$ potential $V$ in the case that the long-range
potential is one-pion exchange, i.e. $V=V^{(0)}$.  (For other research
that bears on the role of one-pion exchange in this channel see
Refs.~\cite{scald,CH98,SF98,KS99,Ge99,Fr99,Ol03,manuel1,manuel3,manuel4,Ti05,Dj07,Ya08}.)
Recent studies, including Refs.~\cite{Hy00,manuel2,En07}, have
examined the impact of higher-order pieces of $V$ on the predictions
that the $\chi$PT approach to nuclear forces makes for phase shifts in
the ${}^1$S$_0$ channel.

Like Ref.~\cite{manuel2}, this paper examines the role of
two-pion-exchange corrections to $V$ in the ${}^1$S$_0$ channel. We
seek to answer two questions. First, is there empirical evidence for
these two-pion exchange contributions, and, concomitantly, what impact
do they have on our understanding of the short-range dynamics in the
$NN$ system?  Second, is there any sense in which $V^{(2)}$ and
$V^{(3)}$ are ``small''? I.e. one (or more) of the expansions proposed
for $NN$ dynamics in Refs.~\cite{wein,Or96,ksw1,Be01} is
converging. The weakness of one-pion exchange in the ${}^1$S$_0$
channel---thanks to the absence of the tensor pieces of the $NN$
potential that generate, e.g. deuteron binding---means that these
questions are of crucial importance in the development of a
quantitative theory of $NN$ scattering in this channel.

In contrast to
Refs.~\cite{Or96,Ep99,Ka97,Ge99,EM02,EM03,Ep05,Ka96,scald,ksw1,ksw2,CH98,SF98,KS99,Fr99,Fl00,Re99,Hy00,Be01,Re03,Ol03,Ti05,Dj07,En07,Ya08},
we do not compare predictions from a treatment of the $NN$ potential
(or amplitude) within a given chiral power counting with data. Instead
we examine the extent to which phase-shift data obtained by the
Nijmegen group~\cite{PWA93,stoks} can be used to make inferences
regarding the $NN$ potentials that are at work for $r < 1/m_\pi$. To
do this we invoke $\chi$PT with only nucleons and pions as explicit
degrees of freedom to obtain $V$ at distances $r \sim
1/m_\pi$~\cite{Or96,Ka97,Ep99}. (The $\chi$PT expansion for $V$
converges quite well in this domain, see, e.g.,
Refs.~\cite{Ka97,Hi04}.)  We then start with the Nijmegen PWA93 phase
shift~\cite{stoks} at a fixed energy and use the long-range potential
at a given order in $\chi$PT to integrate in to a finite distance
$R$. The formalism for this treatment is presented in
Sec.~\ref{sec-outsidein}. Such an ``outside in'' approach, that begins
with phase shifts, and uses a $\chi$PT potential to deduce information
about short-distance dynamics in the ${}^1$S$_0$ channel, was already
pursued in Refs.~\cite{scald,manuel1,manuel3,manuel4}. However, in
these works only the leading-order $\chi$PT potential $V^{(0)}$ was
employed for this purpose. And while the NNLO chiral potential
$V^{(0)} + V^{(2)} + V^{(3)}$ was used to do the integration in
Ref.~\cite{manuel2}, only values $R \ll 1/m_\pi$ were discussed there.

When such an analysis is performed over the range $0 \leq T_{\rm lab}
\leq 200$ MeV it yields information on the energy dependence that the
short-range interaction must have if it is to be used in concert with
the long-range $V$ of $\chi$PT (at a fixed order) to reproduce
data. (This method is very similar to those of the phase-shift
analysis itself, although in Ref.~\cite{stoks} a different long-range
potential was employed.)  Since our effective field theory (EFT) does
not contain explicit Delta degrees of freedom we limit ourselves to
data at fairly low energies: $T_{\rm lab} \leq 200$ MeV.  The results
for the inferred energy dependence due to short-range $NN$ dynamics
are presented in Sec.~\ref{sec-endepres}.  While any energy dependence
due to short-range dynamics is in principle possible, short-range
potentials with coefficients that are natural with respect to the high
scale $1/R$ will lead to smooth, not rapid, energy dependence on the
interval $0 \leq T_{\rm lab} \leq 200$ MeV. As one might expect, if
$R$ is too large, the energy dependence over this entire energy
interval is not smooth, since the $NN$ collision probes details of the
regulator if $p_{\rm cm} =\sqrt{\frac{MT_{\rm lab}}{2}}$ is greater
than or of order $\frac{\pi}{R}$. However, we find that for $R \lsim
1.8$ fm only smooth energy dependence of the short-range potential is
needed, and that this conclusion holds essentially irrespective of the
$\chi$PT long-range potential that governs $NN$ dynamics in the region from
$r=R$ to $r=\infty$.

The second question is whether dynamics in that region can be
understood in perturbation theory. In Section~\ref{sec-pert} we show
that the long-distance pion-exchange interactions derived from
$\chi$PT can indeed be treated in perturbation theory in the region $r
> 1.0$ fm. This result is obtained via a perturbative treatment of the
chiral potential used in the Schr\"odinger equation in
Sec.~\ref{sec-outsidein}. Our perturbation theory preserves the
asymptotic wave function, and hence the phase shifts. In this way we
can examine the extent to which the energy dependence displayed in
Sec.~\ref{sec-endepres} can be understood using perturbation
theory. (A similar long-distance perturbation theory was developed in
Ref.~\cite{manuel2}, but there it was used to integrate the
Schr\"odinger equation to $R \approx 0$, whereas here our focus will
be on finite $R$.)

In Sec.~\ref{sec-slanty} we present one concrete realization of
short-distance physics. We adopt a two-parameter form for the
potential in the region $r \leq R$, and obtain values for the
relevant parameters that generate the various different
energy-dependencies displayed in Sec.~\ref{sec-endepres}.
(We stress, however,
that the results of Sec.~\ref{sec-endepres} for the
boundary condition at $r=R$ do not depend on a particular model of
the physics that is operative
for $r \leq R$.) Finally, in Sec.~\ref{sec-conclusion} we offer some
conclusions and an outlook.

\section{Solving the Schr\"odinger Equation}
\label{sec-outsidein}

We choose to work in co-ordinate space and hence the basic task is
to solve the radial Schr\"odinger equation:
\begin{equation}
{1\over r^2}{d\over dr}\Big(r^2 {d R(r)\over dr}\Big) + {M\over
\hbar^2} (E-V(r)) R(r) = 0.
\end{equation}
Here, $E$ is the c.m. energy of the $NN$ system, and ${M \over 2}$ its
reduced mass. We adopt $M=938.918$~MeV. By the
substitution $u(r) = R(r)/r$ the equation reduces to
\begin{equation}
{d^2 u_k(r)\over dr^2} + \left(k^2 - U(r)\right) u_k(r) = 0,
\label{eq:radialSE}
\end{equation}
where $k^2=\frac{ME}{\hbar^2}$ and $U(r)=\frac{M V(r)}{\hbar^2}$. Our
objective in this section is to solve this equation in the region
$R\leq r$. Because of this, an important feature of our solution is
that we do not use the usual boundary condition $u(0)=0$. Instead, we
invert the problem and use the Nijmegen PWA93~\cite{PWA93} extraction
of the $^1S_0$ phase shifts as input that provides the boundary
condition at $r\rightarrow \infty$:
\begin{equation}
u_k(r)|_{r\rightarrow \infty}=\frac{\sin (kr+\delta(k))}{\sin \delta(k)},
\end{equation}
$\delta(k)$ being the experimental phase-shift corresponding to
c.m.-frame relative momentum $k$. The normalization is done in such a
manner that the asymptotic zero-energy solution is given by $1-r/a$,
$a$ being the scattering length. The purpose of solving the
Schr\"odinger equation in this way is to obtain an energy-dependent
matching condition at any $R$ (which is naively of the order of or
smaller than the range of pion-exchange interaction). This matching
condition is defined by the logarithmic derivative:
\begin{equation}
\gamma(k;R)=\Big[\frac{u'_k(r)}{u_k(r)}\Big]_{r=R}.
\label{eq:gamfull}
\end{equation}
For example,
in the absence of any long-range potential (i.e. the case $V=0$) we
have:
\begin{equation}
\gamma(k;R)=k \cot(kR + \delta(k)).
\label{eq:noLR}
\end{equation}
If we incorporate any long-distance
potential $V$ in the analysis for $r > R$ then the form
(\ref{eq:noLR}) will no longer apply, and in general, we will have
only numerical results for $\gamma(k;R)$.

Regardless, the energy dependence of $\gamma(k;R)$ can be fitted with
a polynomial in $k^2$ as--
\begin{equation}
\gamma(k;R)=\sum_{i} A_i(R)(k^2)^i. \label{eq:gamai}
\end{equation}
The fact that we represent $\gamma(k;R)$ in the form (\ref{eq:gamai})
is guided by the form of the Lagrangian for an EFT of the $NN$
interaction. Because of parity and time-reversal symmetries, only even
powers of $k$ can occur in $NN$ contact interactions.  Such local
interactions (in the sense of quantum field theory), when smeared over
a length scale $R$,  will result in energy dependence that is smooth
with respect to the scale $R$, and so we expect that $A_i(R) \sim
R^{2i-1}$. A lucid description of this philosophy of regulating the
short-range part of a two-body potential can be found
in~\cite{lepage}.

The coefficients in Eq.~(\ref{eq:gamai}) are then manifestations of
the short-range $NN$ interaction in the $^1S_0$ channel. If one wishes
these parameters can in turn be used to construct a model that is a
particular realization of that short-distance physics.  In other
words, once a $\chi$PT potential of a given order had been used to
obtain information on $\gamma(k;R)$, we would use that information
to extract the coefficients of $NN$ contact interactions
$C_0$, $C_2$, $C_4$, etc. This is very much in the spirit of the
Nijmegen PWA93, where Stoks {\it et al.}~\cite{stoks} employed square
wells for $r\leq 1.8$~fm to produce the necessary energy dependence
due to short-distance physics.

But, regardless of what potential generates the coefficients $A_i$ in
Eq.~(\ref{eq:gamai}), the long-range potential ($V(r)$ for $r > R$)
should produce the faster energy dependence in the phase shifts.
Removing this more rapid energy dependence by integrating the wave
function from $r=\infty$ in to $r=R$ and examining $\gamma(k;R)$
allows us to look at the behavior of observables with energy that is
generated by short-distance dynamics. So, although a polynomial form
like (\ref{eq:gamai}) is preferred from the EFT point of view, we are,
at first, agnostic about the form of $\gamma(k;R)$, and merely report
the results obtained from the integrating-in exercise when $\chi$PT
potentials of different orders are adopted in the region $r \geq
R$. But, before we do that, we first give the details of the $\chi$PT
potentials that govern dynamics in that region.

\subsection{The Potential}
\label{sec-pot}

In our analysis we adopt a $\chi$PT expansion for $V(r)$. Since we
will solve the Schr\"odinger equation in co-ordinate
space, we need a co-ordinate-space
representation of the corresponding pion-exchange potentials. We
have adapted the expressions from Ref.~\citep{Ka97}. The potential
corresponding to one-pion exchange is the leading-order part of $V$,
$V^{(0)}$, and  can be expressed as
\begin{equation}
V^{(0)}(r) = (\vec \tau_1 \cdot \vec \tau_2)[\widetilde
W_S^{(1\pi)}(r) (\vec \sigma_1 \cdot \vec \sigma_2) +  \widetilde
W_{T}^{(1\pi)}(r) S_{12}(\hat{r})],
\end{equation}
where $S_{12}(\hat{r})=3 \vec \sigma_1 \cdot \hat{r} \vec \sigma_2 \cdot
\hat{r} - \vec \sigma_1 \cdot \vec\sigma_2$, and
\begin{equation}
\widetilde W_{S}^{(1\pi)}(r) = {g_{\pi N}^2 m_\pi^2 \over 48 \pi
M^2} {e^{- x} \over r} \,\, ,
\end{equation}
\begin{equation}
\widetilde W_T^{(1\pi)}(r) = {g_{\pi N}^2 \over 48 \pi M^2} {e^{-x}
\over r^3} (3+3x +x^2),
\end{equation}
with $x=m_{\pi}r$ and $g_{\pi N}$ the $\pi NN$ coupling
constant. Numerically we choose $\mpi=134.98$~MeV and $g_{\pi
  N}=13.1$~\cite{stoks}.

Similarly, the two-pion exchange potentials which yield $V^{(2)}$ and
$V^{(3)}$ are
\begin{equation}
V^{(2)}(r) = \re (
\widetilde V_{S}(r) (\vec \sigma_1 \cdot \vec \sigma_2)
+ (\vec \tau_1 \cdot \vec \tau_2) \widetilde
W_C(r) + \widetilde V_T(r) S_{12}(\hat{r})),
\end{equation}
for the ``leading-order'' pieces of two-pion exchange, which are
constructed solely out of vertices in ${\cal L}_{\pi N}^{(1)}$, and
\begin{equation}
V^{(3)}(r) = \re (\widetilde V_C(r) + (\vec \tau_1 \cdot
\vec \tau_2) \widetilde
W_{S}^{(2\pi)}(r) (\vec\sigma_1\cdot \vec \sigma_2)
+(\vec \tau_1 \cdot \vec \tau_2 )\widetilde
W_T^{(2\pi)}(r) S_{12}(\hat{r}))
\end{equation}
for ``sub-leading'' two-pion exchange, which involves contributions
from ${\cal L}_{\pi N}^{(2)}$.

The different coefficient functions $\tilde{V}_X(r)$ and
$\tilde{W}_X(r)$---with $X=C$, $S$, $T$ referring
to the central, spin-spin and tensor
components of the potential---are~\cite{Ka97}:
\begin{eqnarray}
\widetilde V_T(r) &=& {g_A^4 m_\pi\over128\pi^3 f_\pi^4 \,r^4}
\left\{-12 x\, K_0(2x) -(15+4x^2)\, K_1(2x)\right\},\label{eq:vt}\\ \widetilde
  W_C(r) &=& {m_\pi \over 128 \pi^3 f_\pi^4\,r^4} \bigg\{\big[1+2
    g_A^2(5+2x^2) -g_A^4 (23+12 x^2) \big] \, K_1(2x) \nonumber \\ 
& &  \qquad+ x \big[1+10 g_A^2 -g_A^4(23+4x^2)\big] \, K_0(2x)\bigg\},\\ 
\widetilde V_{S}(r) &=&
      {g_A^4m_\pi\over32\pi^3f_\pi^4\,r^4} \left\{ 3 x\, K_0(2x)
      +(3+2x^2)\, K_1(2x )\right\},\\
\widetilde  W_{S}^{(2\pi)}(r) &=& {g_A^2\over48 \pi^2 f_\pi^4} {e^{-2x}
        \over r^6} \left\{ c_4 (1+x)(3+3x+2x^2)\right\},\\ 
\widetilde V_C(r)&=&{3g_A^2\over32 \pi^2
        f_\pi^4} {e^{-2x} \over r^6}  \left\{2c_1 x^2(1+x)^2
+  +c_3  (6+12x+10 x^2+4x^3
      +x^4)\right\}, \\ 
\widetilde W_T^{(2\pi)}(r)&=&{g_A^2\over 48
        \pi^2 f_\pi^4} {e^{-2x} \over r^6}\left\{-c_4
(1+x)(3+3x +x^2)\right\}.
\label{eq:wt}\end{eqnarray}

Here, and in the calculations we report on subsequently, we have
omitted the $1/M$ pieces of the potentials derived in
Ref.~\cite{Ka97}.  We find that these contributions have a negligible
effect on the results for $R > 1.4$ fm.  This supports the power
counting adopted in Refs.~\cite{Ep99,Ep05}, where $M \sim \Lambda^2$,
leading to a sub-dominant role for $1/M$ corrections in the $NN$
potential. As $R$ is lowered below 1.0 fm, including $1/M$ corrections
to $V$ leads to marked differences with our results, which raises
questions about the $M \sim \Lambda^2$ power counting in that
domain. However, for $R=1.0$ fm the largest change induced by the
addition of $1/M$ corrections to $V$ is only 15\%. Thus, even there,
they can be regarded as higher order than $V^{(3)}$.

In what follows we have used $g_A=1.29$ to be consistent with our
choice of $g_{\pi N}$ in the LO potential, and $f_{\pi}=92.4$~MeV. For
$c_1$, $c_3$ and $c_4$ we have used two sets of values corresponding
to the low-energy extraction of Rentmeester et. al.~\cite{Re99} and
Entem et. al.~\cite{EM03}. The two sets are tabulated in
Table~\ref{tab1}.
\begin{table}
\caption{Different values of the low-energy constants $c_i$, $i=1,3,4$ (in
GeV$^{-1}$). \label{tab1}}
\begin{center}
\begin{tabular}{ c |c c c}
\tableline
 & $c_1$ & $c_3$ & $c_4$ \\
\tableline
Rentmeester et. al. & -0.76 & -5.08 & 4.70 \\
Entem et. al. & -0.81 & -3.4 & 3.40 \\
\tableline
\end{tabular}
\end{center}
\end{table}

We now need to project out the
potentials that act in the $^1$S$_0$ channel. This removes the tensor components,
leaving us with
\begin{equation}
V^{(0)}(r) = -3 \widetilde W_S^{(1\pi)}(r) , \label{eq:opep}
\end{equation}
which is our leading-order (LO) chiral potential.
Similarly,
\begin{equation}
V^{(2)}(r) = \re (\widetilde W_{C}(r) - 3\widetilde
V_S(r)) , \label{eq:tpep}
\end{equation}
and $V^{(0)}(r)+ V^{(2)}(r)$ is our next-to-leading
order (NLO) chiral potential. And finally,
\begin{equation}
V^{(3)}(r) = \re (\widetilde V_C(r) - 3\widetilde
W_{S}^{(2\pi)}(r)), \label{eq:tpepnl}
\end{equation}
yielding $V^{(0)}(r)+ V^{(2)}(r) + V^{(3)}(r)$ as our
next-to-next-to-leading order (NNLO) chiral potential.

\section{The Energy-Dependent Matching Condition}
\label{sec-endepres}

To examine how the pieces of the long-range $\chi$PT potential
manifest themselves, we solve the Schr\"odinger equation
(\ref{eq:radialSE}) using the LO, NLO, and NNLO potentials
(\ref{eq:opep})--(\ref{eq:tpepnl}). The output of this procedure is
the energy-dependent logarithmic derivative $\gamma(k;R)$, at
different distances $R$.

\begin{figure}[htbp]
\epsfig{figure=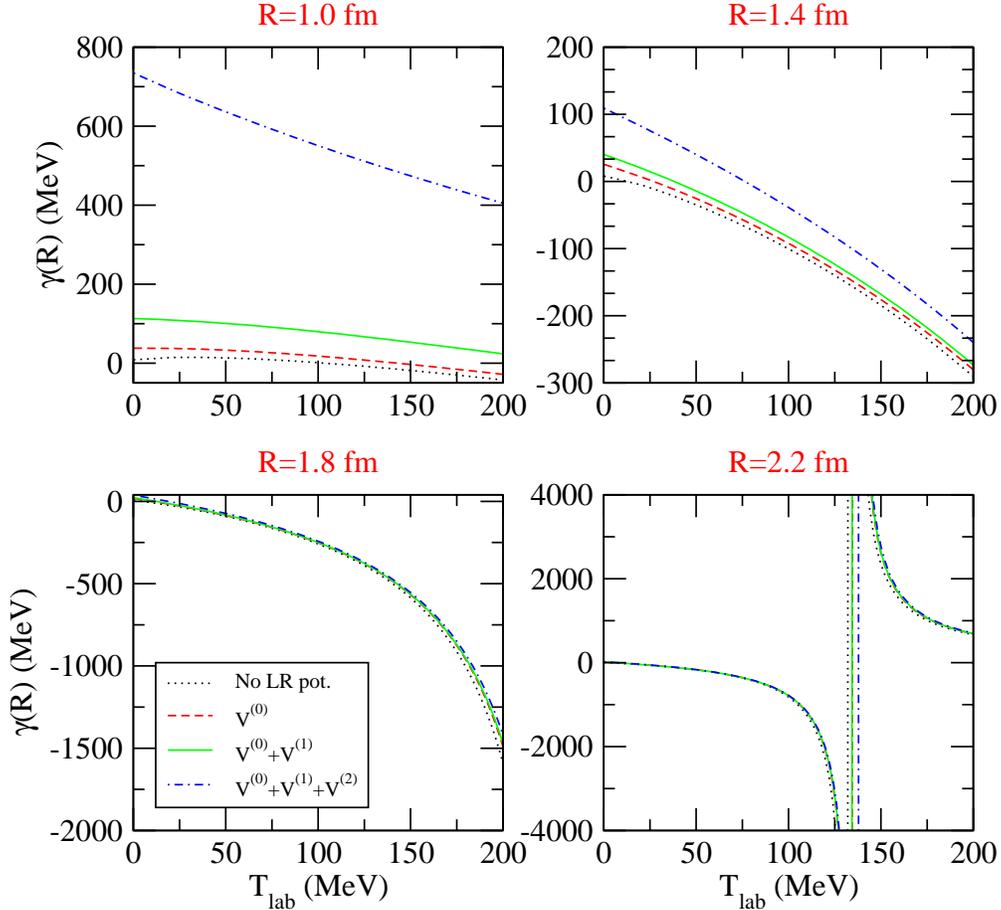, height=5in}
\caption{$\gamma(k;R)$ (MeV) plotted against $T_{lab}$ (MeV) for
different values of $R$. The dotted (black) curve corresponds to
$V(r)=0$, dashed (red) represents $V(r)=V^{(0)}(r)$, solid (green)
represents $V(r)=V^{(0)}(r)+V^{(2)}(r)$ and the dot-dashed (blue)
represents $V(r)=V^{(0)}(r)+V^{(2)}(r)+V^{(3)}(r)$.} \label{fig1}
\end{figure}

The results for $\gamma(k;R)$ are displayed as a function of
lab.~energy $T_{\rm lab}=\frac{2 \hbar^2 k^2}{M}$ for four different
values of $R$ in Fig.~\ref{fig1}. As mentioned before, we present
results for $0 \leq T_{lab} \leq 200$~MeV. The four different panels
correspond to choosing the matching point $R$ to be 1.0 fm, 1.4 fm,
1.8 fm, and 2.2 fm respectively. Each panel itself contains four
different curves, corresponding to different choices for the potential
$V$ in the region $r \geq R$.

Let us first analyze the energy dependence of $\gamma
(k;R)$. Fig.~\ref{fig1} shows that $\gamma (k;R)$ has a smooth
behavior as $T_{\rm lab}$ increases from 0 to 100 MeV. For all $R$'s
examined it could be represented as a polynomial in $k^2$ in this
range. However, at slightly higher energies, $T_{\rm lab}=140$ MeV,
there is a singularity in the energy dependence for $R=2.2$ fm. This
is, in fact, an artifact of the singularity of the cotangent derived
in Eq.~(\ref{eq:noLR}).  The singularity occurs because $k R \approx
\pi$ at these energies, which means that the lab.~energy is high
enough that we are probing details of the dynamics at $r \leq R$. It
is therefore not a surprise that $\gamma(k;R)$ cannot be represented
by a polynomial for $T_{\rm lab} > 140$ MeV. A similar comment
applies to the results for $R=1.8$ fm, although there the
singularity is at $T_{\rm lab}$ a little above 200 MeV.

A smooth, i.e. flatter, energy dependence is highly desirable because
our intention is to use the energy dependence of $\gamma(k;R)$ to
extract information about the short-distance physics per
Eq.~(\ref{eq:gamai}). If we choose $R=2.2$~fm or even $R=1.8$~fm, we
become very limited in the energy-range that can be used for
extraction of the short-distance physics.

Furthermore, the lower panels in Fig.~\ref{fig1} make it clear that
the long-range potentials are not having a significant impact on the
energy dependence of $\gamma(k;R)$ for $R \geq 1.8$ fm. In other
words, no matter what order chiral potential we use, or even if we use
no chiral potential at all, the result for $\gamma(k;R)$ at $R \geq
1.8$ fm looks essentially the same, although there is some effect due
to $V^{(0)}$, i.e. one-pion exchange, in $\gamma(k;1.8$ fm$)$ at
higher energies.

\begin{figure}[t]
\epsfig{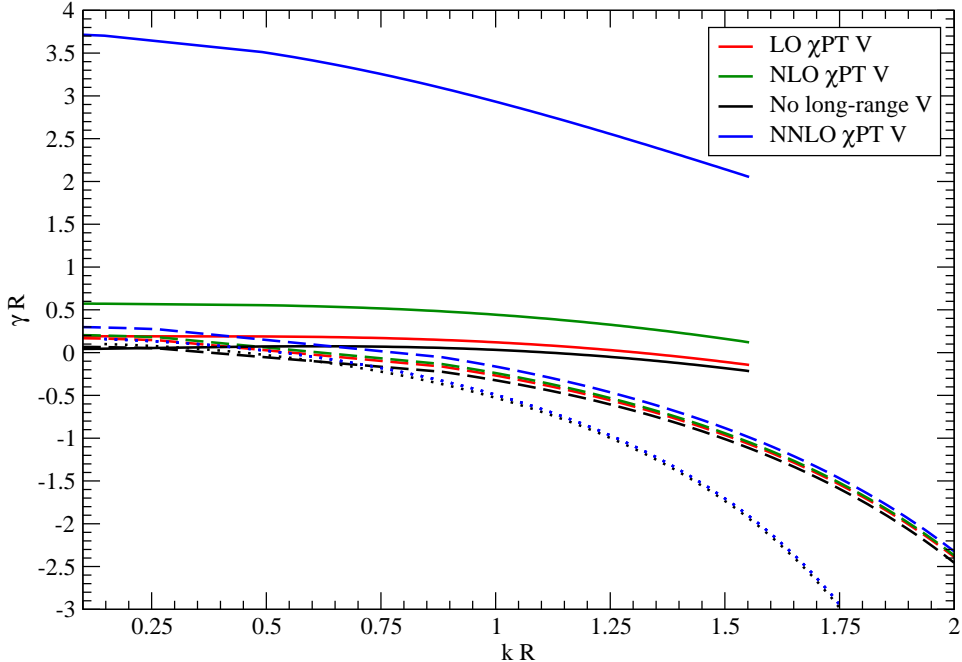} \caption{\label{fig3p} Here
  we have plotted $\gamma R$ vs $kR$, both dimensionless quantities, for
  different $R$s. The dotted curves are for $R=3.0$ fm, the dashed
  curves for $R=1.8$ fm and the solid curves for $R=1.0$ fm. In each
  case the black curve is the result with $V=0$, the red curve
  corresponds to $V=V^{(0)}$, the green curve to $V=V^{(0)}+V^{(2)}$,
  and the blue curve is what is obtained with
  $V=V^{(0)}+V^{(2)}+V^{(3)}$.}
\end{figure}

Both of these phenomena can be seen in one plot if we form the
dimensionless quantity $\gamma R$, and consider the result as a
function of the dimensionless quantity $kR$, as well as dimensionless
numbers formed out of scales present in the $NN$ potential:
\begin{equation}
\gamma R=\gamma R(kR, m_\pi R, \Lambda_{NN} R, \ldots),
\end{equation}
where $\Lambda_{NN}=\frac{16 \pi f_\pi^2}{g_A^2
  M}$~\cite{ksw1,ksw2,BB03} is a scale that sets the strength of the
$NN$ potential at leading order, and the dots indicate the other
scales that will appear if $V^{(2)}$ or $V^{(3)}$ is employed in the
extraction of $\gamma R$ from data.  In Fig.~\ref{fig3p} we have
plotted $\gamma R$ against the rescaled wave number $kR$. The dotted
curves in the figure are generated with $R=3.0$ fm, and we see that
they are all close to the black-dotted curve, which encodes
(\ref{eq:noLR}) at this radius, and so is what is obtained if $V=0$.
In other words, at $R=3.0$ fm essentially all of the ${}^1$S$_0$
phase shift is generated by short-range ($r < 3.0$ fm) dynamics.
This is hardly a surprise given that OPE is the longest-range part
of the $NN$ force, and $m_\pi R=2.1$ at this $R$. But
Fig.~\ref{fig3p} also shows that as we decrease the scale $R$---the
scale that defines the demarcation between ``long-'' and
``short-''distance---there is increased separation between the curve
of Eq.~(\ref{eq:noLR}) and the curves obtained when $V^{(0)}$,
$V^{(0)} + V^{(2)}$, and $V^{(0)} + V^{(2)} + V^{(3)}$ are used for
the analysis. At a separation scale of $R=1.8$ fm pionic effects
generate a larger portion of the overall phase-shift---although the
analysis also shows that this effect comes mostly from one-pion
exchange. But at $R=1.0$ fm we see large effects from pionic
dynamics. At this scale we also see significant differences in the
results for $\gamma R$, depending on what long-range potential is
used for the ``renormalization-group evolution'' from $R=\infty$ to
$R=1.0$ fm. (The formalism for such a renormalization-group analysis
of the phase shift with respect to the scale $R$ was laid out in
Refs.~\cite{manuel1,manuel3,manuel4}.)

Indeed, at the separation scale $R=1.0$ fm, the impact of pion
dynamics in the $NN$ potential on $\gamma R$ is largest for the NNLO
$V$---which includes so-called ``sub-leading'' two-pion
exchange---and the result obtained for $\gamma(k;R)$ does not appear
to be perturbatively close to the (black) no-long-range-potential
curve.  The departure from this no-long-range-potential result is
still visible if one adopts either $V^{(0)}$ or $V^{(0)} + V^{(2)}$
as the long-range potential, but it is nowhere near as large.  This
casts doubt on whether the hierarchy of effects at this separation
scale is still that predicted by $\chi$PT. The issue here, of
course, is that both $V^{(2)}$ and $V^{(3)}$ have stronger
singularities as $r \rightarrow 0$ than does $V^{(0)}$. Thus, as $r$
gets smaller they eventually dominate the evolution of $\gamma(k;R)$
with $R$ and $k$. The $\chi$PT power counting is only a
reliable guide to the relative size of contributions provided $R
\sim 1/m_\pi$.

For these reasons we believe that the separation of long- and
short-distance physics in the ${}^1$S$_0$ channel within a $\chi$PT
framework will be most effective if $1.0~{\rm fm} < R < 1.8~{\rm
  fm}$. If $R > 1.8$ fm there is very little impact from chiral
dynamics on $\gamma(k;R)$. Correspondingly, conversion from the
variable $k R$ back to $T_{\rm lab}$ produces rapid curvature at
disturbingly low values of the laboratory energy. At $R > 1.8$ fm
the computation of $\gamma$ is sensitive to details of the
short-distance dynamics (and not just a few coefficients in the
expansion (\ref{eq:gamai})) for $T_{\rm lab}$ only a little above
100 MeV. In contrast, at $R < 1.0$ fm the short-distance dynamics has less role
in generating the energy dependence of the ${}^1$S$_0$ phase shift;
but the massive attraction generated by $V^{(3)}$ causes problems of
its own, as witnessed by the very large value of $\gamma R$ at $k=0$
in the presence of ``sub-leading two-pion exchange'' that is
inferred at this separation scale.

We now take the results of Figs.~\ref{fig1} and \ref{fig3p} and
extract the coefficients $A_0$ and $A_1$ in the expansion
(\ref{eq:gamai}) of $\gamma(k;R)$ for different choices of the
long-range potential $V$. These results will be used in
Section~\ref{sec-slanty} when we attempt to implement specific
short-distance potentials that reproduce the low-energy behavior of
$\gamma(k;R)$.  The results for $\gamma(k;R)$ obtained at two
different separation scales $R$ in the range $1.0~{\rm fm} < R <
1.8~{\rm fm}$ are presented in Table~\ref{tab2}. (For the results
tabulated we have have used four terms in the expansion
(\ref{eq:gamai}) to convince ourselves that the coefficients $A_2$,
$A_3$, etc. are indeed much smaller compared to either $A_0$
or$A_1$. For example, $A_2\sim 10^{-3}$ fm$^3$ and $A_3\sim 10^{-5}$
fm$^5$ for $R=1.4$ fm.) The numbers in Table~\ref{tab2} reinforce
the conclusion that the next-to-leading-order two-pion exchange
potential has the largest effect, with a pronounced impact on $A_0$
and $A_1$ for $R=1.0$~fm. We note that $A_2$ and $A_3$ are small,
and the values of $A_0$ and $A_1$ shown are natural with respect to
the scale $R$ and the underlying scale of $\chi$PT, $\Lambda_{\chi
\rm SB}$.

\begin{table}[htb]
\caption{The coefficients $A_0$ (in fm$^{-1}$) and $A_1$ (in fm) for
$R=1.0$~fm and $R=1.4$~fm for different choices of long-range
potential. \label{tab2}}
\begin{center}
\begin{tabular}{ | c | c  c | c  c |}
\tableline
Potential & \multicolumn{2}{c}{R=1.0 fm} \vline& \multicolumn{2}{c}{R=1.4 fm} \vline \\
 & $A_0$& $A_1$& $A_0$ & $A_1$ \\
\tableline
No LR pot & 0.042 & 0.328 & 0.045 & -0.284 \\
$V^{(0)}$ & 0.192 & 0.008 & 0.130 & -0.37 \\
$V^{(0)}+V^{(2)}$ & 0.403 & -0.044 & 0.178 & -0.39 \\
$V^{(0)}+V^{(2)}+V^{(3)}$ & 3.101 & -0.648 & 0.496 & -0.518 \\
\tableline
\end{tabular}
\end{center}
\end{table}

%\begin{figure}[htbp]
%\epsfig{figure=071001_diffRs_noM.eps, height=2.5in}
%\caption{\label{fig2} $\gamma(k;R)$ (MeV) vs. $T_{lab}$ (MeV) for
%three different values of $R$ with
%$V(r)=V^{(0)}(r)+V^{(2)}(r)+V^{(3)}(r)$.}
%\end{figure}

\begin{figure}[htbp]
\epsfig{figure=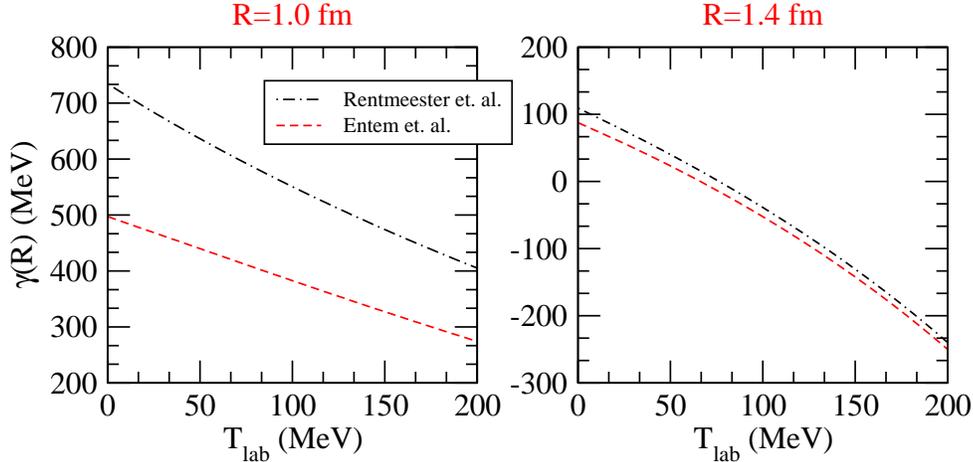, height=2.5in}
\caption{\label{fig3} Effect of different sets of $c_i$'s from
Table~\ref{tab1} on the energy-dependent matching condition at
$R$$=1.0$ fm and $1.4$ fm with
$V(r)=V^{(0)}(r)+V^{(2)}(r)+V^{(3)}(r)$. }
\end{figure}

Finally, if sub-leading two-pion-exchange effects are so critical one
must be cautious about the set of low-energy constants chosen for
their evaluation.  Fig.~\ref{fig3} shows the effect of using different
values of these $c_i$'s in our calculation. The two choices are those
tabulated in Table~\ref{tab1}. It is obvious that the Nijmegen set of
$c_i$'s produces a stronger effect in $\gamma (k;R)$, which is not
surprising given that the combination $c_3 - 2c_4$ sets the size of
the leading singularity in $V^{(3)}$ in this channel, and this
combination is much larger for the $c_i$'s adopted in
Ref.~\cite{Re99}. It is interesting to note that both choices lead to
a similar shape for $\gamma(k;R)$, although the magnitude is larger in
the case of the Rentmeester {\it et al.} choice. Presumably the fact
that $\gamma(k;R)$ is smooth in both panels allows either choice of
$c_i$'s to yield a reasonable fit for $NN$ scattering data that is
sensitive to the ${}^1$S$_0$ phase shift. It is just that the
separation between the long-distance dynamics in $V^{(3)}$ and the
short-distance dynamics encoded in $\gamma R$ will differ, depending
on the value of the $c_i$'s that is adopted. From this point on all of
the results that we present use the Nijmegen set of $c_i$'s.

\section{Are pion exchanges perturbative?}
\label{sec-pert}

In the previous section we showed that the logarithmic derivative
$\gamma(k;R)$ is a logical way to look at the short-distance parts of
the $NN$ interaction. In this section we turn to the question of
whether the long-range part of $V$ can be treated
perturbatively. To do this we analyze how close a perturbative
treatment of the $\chi$PT potential is to the full solution
of the Schr\"odinger equation in the region $r > R$.

Our objective is still to solve the Schr\"odinger equation:
\begin{equation}
{d^2 u(r)\over dr^2} + (k^2-U(r)) u(r) = 0,
\label{eq:pertSE}
\end{equation}
in the region $r>R$. Here $U(r)=\frac{M V(r)}{\hbar^2}$ is obtained
from the potential in this region of space, which in our case is the
LO, NLO or the NNLO potentials generated in $\chi$PT.  In order to
build up the solution, let us first rewrite Eq.(\ref{eq:pertSE}) as:
\begin{equation}
-{d^2 u_k(r)\over dr^2} + \alpha U(r) u_k(r) = k^2
u_k(r), \label{eq:pertSE2}
\end{equation}
where $\alpha$ is a parameter that is used to describe the
perturbative expansion in powers of $V$ (note that this is not the
same as the $\chi$PT expansion for $V$ itself that was described in
Sec.~\ref{sec-intro}), and the subscript $k$ refers to the wave number
of interest. The full solution can now be written as a power series in
$\alpha$ as:
\begin{equation}
u_k(r)=\sum_{n=0}^{\infty} \alpha^n u_k^{(n)}(r)= u_k^{(0)}(r)+
\alpha u_k^{(1)}(r)+ \alpha^2 u_k^{(2)}(r)+\ldots ,
\label{eq:pertexp}
\end{equation}
where, $u^{(0)}(r)$ is the zeroth-order solution (in the absence of
$V(r)$), $u^{(1)}(r)$ is the first-order-in-$V$ correction, $u^{(2)}(r)$ is
the second-order correction, and so on.

Substituting Eq.~(\ref{eq:pertexp}) into Eq.~(\ref{eq:pertSE2})
and equating powers of $\alpha$ we get:
\begin{eqnarray}
{d^2 \over dr^2} u^{(0)}(r) + k^2 u_k^{(0)}(r)&=& 0, \label{eq:zerothorder} \\
{d^2 \over dr^2} u_k^{(n+1)}(r) + k^2 u_k^{(n+1)}(r)&=& U(r)
u_k^{(n)}(r); \qquad n \geq 0.\label{eq:pertsolk}
\end{eqnarray}

The solution of Eq.~(\ref{eq:zerothorder}) that reproduces the PWA93
${}^1$S$_0$ phase shift $\delta(k)$ is:
\begin{equation}
u_k^{(0)}(r) = {\sin (kr+\delta(k))\over \sin \delta(k)}. \label{u0}
\end{equation}
Equations (\ref{u0}) and (\ref{eq:pertsolk}) can now be used, together
with the Green's function technique, to calculate the corrections to
$u$ at first order and second order in perturbation theory.  The
Green's function for Eq.~(\ref{eq:pertsolk}) that preserves the form
(\ref{u0}) as $r \rightarrow \infty$ in the full solution $u_k(r)$ is:
\begin{equation}
G(r,r';k) = \left\{\begin{array}{cl}
\frac{\sin k(r-r')}{k} & \mbox{if $r < r'$},\\
0 & \mbox{if $r > r'$}.
\end{array} \right.
\end{equation}
Therefore the first-order correction to the wavefunction can be expressed as:
\begin{eqnarray}
u_k^{(1)}(r) &=&  \int_r^{\infty}G(r,r';k) U(r') u_k^{(0)}(r') dr'
\nonumber \\
 &=&  {M \over \hbar^2}\int_r^{\infty}{\sin k(r-r') \over k} V(r')
          {\sin (kr' +\delta(k)) \over \sin \delta(k)} \, dr'.
\label{eq:FOPT}
\end{eqnarray}

The second-order correction to the wavefunction is then calculable as:
\begin{equation}
u_k^{(2)}(r) = \Big({M\over \hbar^2}\Big) \int_r^{\infty}G(r,r';k) V(r')
u_k^{(1)}(r') \, dr'.
\label{eq:SOPT}
\end{equation}
Comparison of results from Eqs.~(\ref{eq:SOPT}) and (\ref{eq:FOPT})
will allow us to assess the convergence of our ``long-distance
perturbation theory''. This perturbation theory was set up in
Refs.~\cite{CH98,manuel2}, and was used to discuss the convergence of
the chiral expansion. In contrast to Ref.~\cite{manuel2},
which integrated Eqs.~(\ref{eq:FOPT}) and (\ref{eq:SOPT}) to $R=0$,
we will only integrate them to some finite $R$. Hence, we once again
examine the logarithmic derivative at a radius $R$, only this time we
define a version that can be computed using long-distance perturbation
theory:
\begin{equation}
\gamma^{(n)}(k;R) \equiv \left[\frac{(u_k^{(0)}+u_k^{(1)}+\ldots
    u_k^{(n)})'(r)}{(u_k^{(0)}+u_k^{(1)}+\ldots u_k^{(n)})(r)}\right]_{r=R}
\label{eq:pertgam}
\end{equation}
Strictly speaking this $\gamma^{(n)}$ includes effects of $V$ which
are of an order higher than $n$, but it has the advantage that it is
straightforward to compute. It also results in smooth energy dependence
when the limit $R
\rightarrow 0$ is taken~\cite{manuel4,En07}.

\begin{figure}[htbp]
\epsfig{figure=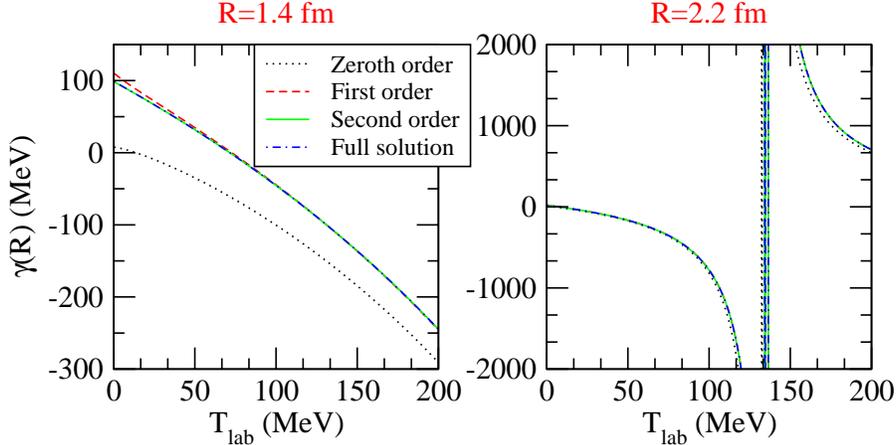, height=2.5in}
\caption{\label{fig4} $\gamma(k;R)$(MeV) vs. $T_{lab}$(MeV)
calculated in perturbation theory and compared to the full solution.
The dotted (black) curve is the zeroth order solution, the dashed
(red) curve is the first order perturbation theory result, the solid
(green) curve is the second order result and the dot-dashed (blue)
curve is from the full solution of the Schr\"odinger equation.}
\end{figure}

To test the usefulness of Eq.~(\ref{eq:pertgam}) we choose the
strongest potential, which, for distances $R < 1.8$ fm, we know to be
the NNLO potential $V(r)=V^{(0)}(r)+V^{(2)}(r)+V^{(3)}(r)$, and
compare the results with those of the previous section.
We again adopt the $c_i$'s of Rentmeester {\it et
  al.}~\cite{Re99}. The results are plotted in Fig.~\ref{fig4}. The
dotted (black) curve is the zeroth-order solution, which corresponds
to using the free Schr\"odinger equation for the integration from
$r=\infty$ to $r=R$.  This, then, is the same as the ``No-long-range
potential'' case of the previous section. Meanwhile, the dashed (red)
curve is the first-order perturbation theory result, the solid (green)
curve is the second-order result, and the dot-dashed (blue) curve is
from the full solution of the Schr\"odinger equation obtained in
Sec.~\ref{sec-endepres}.

The right panel of Fig.~\ref{fig4} shows that---as one might expect
given the results of that section---at $R=2.2$ fm this potential has
almost no impact on $\gamma(k;R)$, and the result for any finite
order of perturbation theory is the same. The left panel of
Fig.~\ref{fig4} reiterates that at $R=1.4$ fm the $\gamma(k;R)$
results with the NNLO potential are very different from the
``No-long-range potential case''. But it also shows that this
difference is almost entirely due to the first-order
perturbation-theory correction. There is a 10\% difference between
the first-order result, $\gamma^{(1)}$ and the full $\gamma$ at $k
\approx 0$, but this 10\% difference has almost completely
disappeared by $T_{\rm lab}=100$~MeV: perturbation theory works
better at higher energies.  And even at the lower energies where
there is some noticeable deviation, the inclusion of the
second-order correction brings us to within better than 1\% of the
full result for $\gamma(k;R)$.  This success of perturbation theory
implies that the coefficients $A_0$ and $A_1$ defined in
Eq.~(\ref{eq:gamai}) can instead be extracted using the
perturbative formulae (\ref{u0}), (\ref{eq:FOPT}), and
(\ref{eq:SOPT})---rather than from the full Schr\"odinger equation
solution---without introducing undue error.

\begin{table}[htb]
\caption{The coefficients $A_0$ (in fm$^{-1}$) and $A_1$ (in fm) for
  $R=1.4$~fm extracted using Eq.~(\ref{eq:pertgam}). The second and
  third (fourth and fifth) columns show the $A_0$ and $A_1$
  corresponding to $\gamma^{(1)}$ ($\gamma^{(2)}$). The sixth and
  seventh columns repeat the result of Table~\ref{tab2p}, for
  comparison.  \label{tab2p}}
\begin{center}
\begin{tabular}{ | c | c  c | c  c |c c|}
\tableline
Potential & \multicolumn{2}{c}{$\gamma^{(1)}$} \vline&
\multicolumn{2}{c}{$\gamma^{(2)}$}  \vline&
\multicolumn{2}{c}{Full soln.} \vline\\
 & $A_0$& $A_1$& $A_0$ & $A_1$ & $A_0$ & $A_1$\\
\tableline
$V^{(0)}$ & 0.130 & -0.369 & 0.130 & -0.369  & 0.130 & -0.37 \\
$V^{(0)}+V^{(2)}$ & 0.18 & -0.401 & 0.178 & -0.391 & 0.178 & -0.39 \\
$V^{(0)}+V^{(2)}+V^{(3)}$ & 0.569 & -0.624 & 0.516 & -0.534 & 0.496 & -0.518 \\
\tableline
\end{tabular}
\end{center}
\end{table}

In Table~\ref{tab2p} we do exactly that, using the perturbative formula
(\ref{eq:pertgam}) to determine $A_0$ and $A_1$ for the case $R=1.4$
fm.  The numbers agree well with those in Table~\ref{tab2} for the
``full solution''. A first-order-perturbation-theory extraction is
sufficient for $V(r)=V^{(0)}$ and $V(r)=V^{(0)}+V^{(2)}$. For
$V(r)=V^{(0)}+V^{(2)}+V^{(3)}$ a second-order-perturbation-theory
calculation does a noticeably better job.

Since the potentials get stronger as $r \rightarrow 0$ the convergence
of perturbation theory is rather slow for $R=1.0$ fm.  We find that
for $R=1.0$~fm, $\gamma^{(1)}(k;R)$ overshoots $\gamma(k;R)$ by almost
a factor of two at $k=0$. The second-order correction brings
$\gamma^{(2)}(k;R)$ to within 10\% of the ``exact'' result at $k=0$,
and the agreement is better than 5\% at $T_{\rm lab}=100$ MeV. So
perturbation theory formally converges for $R=1.0$ fm, but the
``chiral'' nature of the perturbation theory is very much in question,
since---as emphasized in Sec.~\ref{sec-endepres}---the effects of
$V^{(3)}$ are much larger than those of $V^{(0)}$ in this region.

We conclude that the impact of the pion-exchange potentials generated
in $\chi$PT can be calculated in perturbation theory,
provided that the region of $r$ where that perturbation theory is applied
is chosen judiciously. It is possible that a more sophisticated
perturbation theory works even if $R \leq 1.0$ fm, but what is
already clear from these results is that standard ``long-distance''
perturbation theory is applicable and useful in the domain $R >
1.0$ fm.

\section{Short-Distance Parameters and Potential Wells}
\label{sec-slanty}

We now change gears and discuss how the information gleaned from the
energy-dependent matching condition $\gamma(k;R)$ can be used to
extract useful results regarding the short-distance physics that is
operative at $r<R$.

The coefficients defined in Eq.~(\ref{eq:gamai}) and listed in
Table~\ref{tab2} for different choices of $R$ and $V(r)$ are
numbers. But, they can be used to build a short-distance potential
that is, in effect, a short-distance regulator for our $NN$
interaction. Our results suggest that two terms are sufficient to
ensure convergence in the expansion in Eq.~(\ref{eq:gamai}). Thus, any
short-range potential we build having two parameters can be correlated
to the coefficients $A_0$ and $A_1$. We have chosen to design our
short-range potential to be a well with a bottom that has a slope so
that
\begin{equation}
V_{SR}(r)=-V_0 + \mu r, \quad \mbox{for~ $r\leq R$}. \label{eq:vsr}
\end{equation}
Here, $V_0$ (the strength of the short-distance potential at $r=0$)
and the slope $\mu$ are the two parameters of our regulator which we
denote as short-distance parameters (SDPs). In order to extract these
two parameters we solve the Schr\"odinger equation for $r\leq R$ such
that the following condition is satisfied:
\begin{equation}
\Big[\frac{du_{in}'(r)}{u_{in}(r)}\Big]_{r=R}=\gamma (k;R).
\label{eq:vin}
\end{equation}
The motivation in choosing the short-range potential in this form is
that through an appropriate change of variable from $r$ to $x$,
where
\begin{equation}
x=\left( \frac{M \mu}{\hbar^2}\right)^{1 \over 3} \left[r - \frac{E + V_0}{\mu}\right],
\end{equation}
the Schr\"odinger equation for $r < R$ can be reduced to the Airy equation:
\begin{equation}
{d^2 u(x)\over dx^2} - x u(x) = 0. \label{eq:Airy}
\end{equation}
The solution to
Eq.~(\ref{eq:Airy}) is a linear combination of Airy functions:
\begin{equation}
u(x)= a_1 Ai(x)+a_2 Bi(x).
\end{equation}
The Airy function $Bi(x)$ diverges at $x\rightarrow \infty$, but
we are interested in the region $0<r<R$, and so must keep both
solutions. $a_1$ and $a_2$ are arbitrary constants that are
evaluated from boundary conditions $u(0)=0$ and $u'(0)=1$. Thus
changing the variable back to $r$ we obtain--
\begin{equation}
u(r)=\pi\left( \frac{\hbar^2}{M \mu} \right)^{1\over 3}\left[ -Bi
(x_0) Ai\left( \tilde{r} +
x_0\right) + Ai(x_0) Bi \left( \tilde{r} + x_0\right) \right]
\end{equation}
where, $x_0=-(\frac{M \mu}{\hbar^2})^{1\over 3}\frac{E+V_0}{\mu}$, $\tilde{r}=
(\frac{M \mu}{\hbar^2})^{1\over 3} r$.

From this solution at $r \leq R$ we can easily obtain its logarithmic
derivative at $r=R$:
\begin{equation}
\frac{u'(R)}{u(R)}=\left( \frac{\hbar^2}{M \mu} \right)^{1\over 3}
\frac{Bi (x_0) Ai' ( y_0) -  Ai(x_0) Bi' (y_0)}
{Bi (x_0) Ai( y_0)  -  Ai(x_0) Bi (y_0)},
\end{equation}
with $y_0=(\frac{M \mu}{\hbar^2})^{1\over 3} [R - \frac{E +
    V_0}{\mu}]$.  This logarithmic derivative is a function of our
SDPs, $V_0$ and $\mu$. It should be equal to the logarithmic
derivative $\gamma (k;R)$ computed in
Sec.~\ref{sec-endepres}. Hence, using the lowest two terms in the
Taylor expansion of that $\gamma(k;R)$, $A_0$ and $A_1$, we can extract
$V_0$ and $\mu$.

\begin{table}[htb]
\caption{The short-distance parameters $\mu$ (in MeV fm$^{-1}$) and
  $V_0$ (in MeV) for $R=1.4$~fm with different choices of
  long-distance potential. (Note that in our convention $V_0 < 0$
  corresponds to a repulsive potential.) The second and third columns
  give the results that match the Taylor-series coefficients $A_0$ and
  $A_1$ of $\gamma(k;R)$ defined in Eq.~(\ref{eq:gamfull}). The fourth
  and fifth (sixth and seventh) columns provide SDPs that match the
  $A_0$ and $A_1$ for $\gamma^{(1)}$ ($\gamma^{(2)}$) of
  Eq.~(\ref{eq:pertgam}).} \label{tab3}
\begin{center}
\begin{tabular}{ | c | c c | c c | c c |}
\tableline
Potential & \multicolumn{2}{c}{Full soln.} \vline& \multicolumn{2}{c}{$\gamma^{(1)}$} \vline& \multicolumn{2}{c}{$\gamma^{(2)}$}\vline \\
  & $\mu$ & $V_0$ & $\mu$ & $V_0$ & $\mu$ & $V_0$\\
\tableline
No LR pot & 9.85 & 59.35 & 9.85 & 59.35 & 9.85 & 59.35  \\
$V^{(0)}(r)$ & -54.5 & -10.68 & -54.5 & -10.56 & -54.5 & -10.56 \\
$V^{(0)}(r)+V^{(2)}(r)$ & -56.0 & -17.39 & -55.5 & -18.23 & -56.0 & -16.94 \\
$V^{(0)}(r)+V^{(2)}(r)+V^{(3)}(r)$ & -65.0 & -54.4 & -64.0 & -55.65 & -65.0 & -51.1 \\
\tableline
\end{tabular}
\end{center}
\end{table}

To demonstrate, we have calculated the short-distance
parameters, $V_0$ and $\mu$, in this way for $R=1.4$~fm. The results are
given in Table~\ref{tab3}. Results for a different two-parameter
short-distance potential and the case of $V=0$ and $V=V^{(0)}$ can be
found in Ref.~\cite{scald}.

We had commented earlier in Sec.~\ref{sec-pert} that $A_0$ and
$A_1$ can be extracted from the matching condition calculated in
perturbation theory.  Consequently, one can then proceed to extract
$\mu$ and $V_0$ from these ``perturbative'' $A_0$ and $A_1$. The
results are shown in the third to sixth column of Table~\ref{tab3} and
are gratifyingly close to those obtained from the full Schr\"odinger
equation solution of Sec.~\ref{sec-endepres}.

The results of Table~\ref{tab3} show that a nice quantitative
description of the $NN$ interaction in the $^1$S$_0$ channel up to
$T_{\rm lab}=200$ MeV can be obtained by using perturbation theory for
$r\geq R$ to calculate $\gamma(k;R)$, then using that information to
determine the parameters present in a simple short-distance potential
such as (\ref{eq:vsr}).

\section{Conclusion}
\label{sec-conclusion}

Our analysis of $NN$ scattering in the ${}^1$S$_0$ channel is similar
in its philosophy to that advocated by Lepage~\cite{lepage}, in that
we separate the potential into a long-distance part---determined by
$\chi$PT---and a short-distance part, which can be parameterized in a
variety of ways without affecting the predictions for $NN$
scattering. Indeed, we have gone further than Lepage, and argued that
the energy-dependent logarithmic derivative at $R$, $\gamma(k;R)$, is a
convenient way to summarize information about the impact of physics
that is short-range with respect to the scale $R$. This information on
the short-range $NN$ dynamics can be obtained from the $NN$ phase
shifts if we know the long-range potential.  Combining $\chi$PT
long-range potentials (computed to different orders in the chiral
expansion) with the experimental phase-shifts from the Nijmegen PWA93
we obtained $\gamma(k;R)$ for a range of scales $R$.

For the extraction of model-independent information on short-distance
physics it is desirable that $\gamma(k;R)$ be a smooth function of
$k$. This requirement limits the energy range within which one can
operate if $R$ is chosen to be greater than $1.8$~fm. For values of
$R$ in the range from about 1.0 fm to 1.8 fm $\gamma(k;R)$ can be well
described by a polynomial in $k^2$ over the entire range $0 \leq
T_{\rm lab} \leq 200$ MeV. The coefficients of the terms in this
polynomial represent the effect of the short-distance $NN$
interaction. We have found that for $1.0~{\rm fm} < R < 1.8~{\rm fm}$
the first two terms of this polynomial expansion are sufficient to
ensure a good representation of the energy dependence of $\gamma$.
So, in this range of $R$, details of the short-distance potential are
not important: its effects can be summarized in $A_0$
and $A_1$.

The long-range potential $V$ determines the evolution of $\gamma(k;R)$
with the separation scale $R$ (see Refs.~\cite{manuel3,manuel4} for a
derivation of the renormalization-group equations associated with this
evolution).  The scale $R$ in our analysis plays a similar role to the
scale $\Lambda$ in the potential $V_{{\rm low} k}$~\cite{Bo03}. In
that approach $\Lambda$ is a cutoff that separates the momentum-space
states explicitly included in the Hilbert space from those states
whose effects are encoded in $V$.  There is then evolution of the $NN$
potential $V$ with $\Lambda$ so that S-matrix elements (which are
equivalent to phase shifts) stay the same as the momentum-space cutoff
is changed. In our work the short-distance part of the potential also
evolves in such a way that the ${}^1$S$_0$ phase shifts always agree
with those from the Nijmegen PWA93. This evolution can be traced over
a range from $R=\infty$ to $R=1.0$ fm. The fact that, at the lower end
of this range, details of the short-distance potential do not play a
key role in describing $NN$ phase shifts below $T_{\rm lab}=200$ MeV,
is presumably a corollary of the existence of a``universal'' $V_{{\rm
    low} k}$ if the scale $\Lambda$ is lowered to 2 fm$^{-1}$.

In the energy range $0 \leq T_{\rm lab} \leq 200$ MeV we find that the
one-pion exchange interaction is the dominant effect in this evolution
for $R \geq 1.8$ fm, but its impact on $\gamma(k;R)$ is not sizeable.
We obtain a larger shift in $\gamma(k;R)$ compared to its value in the
absence of any long-range potential if we consider sub-leading pieces
of the $\chi$PT $V$. In particular, if we compute with $V^{(3)}$, and
consider $R \leq 1.4$ fm, we see a significant impact of
pion-exchange dynamics on the $\gamma(k;R)$ inferred from the $NN$
phase shifts.

But, even in this case, as long as we maintain $R > 1.0$ fm, we find
that the evolution of $\gamma(k;R)$ due to long-distance effects can
be understood in perturbation theory. Hence we can use perturbation
theory in the $\chi$PT potentials in the region $1.0~{\rm fm} < R <
\infty$ to connect experimental data to information on the
short-distance dynamics in the region $R < 1.0$ fm.  The
short-distance part then contains non-perturbative dynamics, but
these effects are summarized in a few coefficients in the polynomial
expansion of the energy-dependent matching condition. Those
coefficients can in turn be matched to an explicit realization of
short-distance physics, e.g. the one discussed in
Sec.~\ref{sec-slanty}.

This represents an updated version of the proposal first made by
Kaplan, Savage, and Wise in Refs.~\cite{ksw1,ksw2}: that the contact
interaction parameterizing short-distance $NN$ dynamics should be
promoted to order $Q^{-1}$ so that it is a leading-order effect,
with pion exchanges then being sub-leading. This proposal was
examined in detail for $NN$ scattering in
Refs.~\cite{Fl00,Be01}, and was found to work well in the ${}^1$S$_0$
channel as far as the pion potential $V^{(0)}$ was concerned. We
find that $V^{(2)}$ can also be treated in perturbation theory, and
its effects are smaller than $V^{(0)}$ for $r > 1.0$ fm. The NNLO
piece of $V$, $V^{(3)}$, can also be treated in perturbation theory,
but only if it is regulated at a scale of order, or well below, the
chiral-symmetry breaking scale $\Lambda_{\chi {\rm SB}}$. The large
size of corrections due to $V^{(3)}$ raises the question of whether
the chiral expansion for $V$ is behaving as decreed by the $\chi$PT
power counting.

In particular, some authors have suggested that an expansion in which
the Delta-nucleon mass difference, $\Delta$, is treated as a light
scale, might have better convergence
properties~\cite{JM91,Or96,He97,Ka98,Pa03,Pa05,Kr07}. Our results
support this view---at least if one takes $R$ to be small enough that
chiral dynamics plays a significant role. More recently Robillota has
pointed out that the poor convergence of $\chi$PT without an explicit
Delta for the scalar-isoscalar piece of the $NN$ potential is related
to issues with the description of the nucleon's scalar form
factor~\cite{Ro07}. He observes that in that case too, contributions
from Delta loops exceed those from nucleon loops once one considers
distances $r$ of 1 fm or less.

We close with some possible implications of our results for future
partial-wave analyses (PWAs). Here we have used the Nijmegen-extracted
phase shifts~\cite{PWA93} and our separation of $NN$ dynamics into
long-range and short-range parts is similar to that adopted in the
Stoks et. al. 1993 PWA~\cite{stoks}. However, in Ref.~\cite{stoks}
an upgraded version of the Nijm78 potential
was used to describe the long-distance dynamics. Minimally, this
paper shows that in the ${}^1$S$_0$ channel we can use $\chi$PT
potentials and still get sensible energy dependence in our matching
condition which encodes the short-distance $NN$ dynamics. We have
shown that the region $1.0 \leq R \leq 1.8$ fm yields best results
in this regard. This is in accord with the more recent PWA for which
details are as yet unpublished~\cite{Re99,Re03,ectstartalk}. But our results
suggest that---at least in the ${}^1$S$_0$ channel---future PWAs
using these values of $R$ could employ perturbation theory to
compute the effects of the long-distance potentials obtained in
$\chi$PT.  It might also be possible to lower the separation scale
$R$ below 1.0 fm and encode the smooth energy dependence due to
dynamics at $r < R$ in energy-dependent square wells, but then the
parameters obtained for these wells would be very different
depending on whether the LO or NNLO $NN$ potential were used---or
even presumably, if different sets of $c_i$'s were used in computing
$V^{(3)}$. Extensions of the methods laid out here to other partial
waves are straightforward and would facilitate such a future
EFT-based partial-wave analysis.

\section*{Acknowledgments}
This work was carried out under grant DE-FG02-93ER40756 of the US-DOE
(DS, DP, EM) and by NSF grant PHY-0645468 (DS). DP is grateful to
Silas Beane and Dick Furnstahl for useful discussions on the topics
covered here. We also thank Manuel Pav\'on Valderrama, Enrique Ruiz
Arriola, and Matthias Schindler for stimulating conversations as well as
their careful reading of, and informative comments on, this
manuscript.

\end{document}